\documentclass[useAMS,usenatbib]{mn2e}
\usepackage{times,graphicx,amsmath,amsfonts,amssymb,aas_macros,color}
\usepackage{epstopdf}

\newcommand{\beq}{\begin{equation}}
\newcommand{\eeq}{\end{equation}}
\def\beqa{\begin{eqnarray}}
\def\eeqa{\end{eqnarray}}
\def\hmpc{h^{-1}\,{\rm Mpc}}

\def\difd{\textrm{d}}

\title[Skewness as a probe of BAOs]{Skewness as a probe of Baryon Acoustic Oscillations}
\author[Roman Juszkiewicz, Wojciech A. Hellwing, Rien van de Weygaert]{Roman Juszkiewicz$^{1,2}$, Wojciech A. Hellwing$^{1,3,4}$\thanks{e-mail: pchela@icm.edu.pl}, Rien van de Weygaert$^{5}$\thanks{e-mail: weygaert@astro.rug.nl}\\
$^{1}$Institute of Astronomy, University of Zielona G\'ora, ul. Lubuska 2, Zielona G\'ora, Poland\\
$^{2}$Nicolaus Copernicus Astronomical Center, ul. Bartycka 18, Warsaw, 00-716, Poland\\
$^{3}$Institute for Computational Cosmology, Department of Physics, Durham University, Science Laboratories, South Rd, Durham DH1 3LE, UK\\
$^{4}$Interdisciplinary Centre for Mathematical and Computational Modeling (ICM), University of Warsaw, ul. Pawi\'nskiego 5a, Warsaw, Poland\\
$^{5}$Kapteyn Astronomical Institute, University of Groningen, P.O. Box 800, 9725LB Groningen, the Netherlands
}
\begin{document}

\date{Accepted XXXX . Received XXXX; in original form XXXX}

\pagerange{\pageref{firstpage}--\pageref{lastpage}} \pubyear{2012}

\maketitle

\label{firstpage}

\begin{abstract}
In this study we show that the skewness $S_3$ of the cosmic density field contains a significant 
and potentially detectable and clean imprint of Baryonic Acoustic Oscillations. Although the 
BAO signal in the skewness has a lower amplitude than second order measures like the two-point correlation 
function and power spectrum, it has the advantage of a considerably lower sensitivity to systematic 
influences. Because it lacks a direct dependence on bias if this concerns simple linear bias, skewness 
will be considerably less beset by uncertainties due to galaxy bias. Also, it has a weaker sensitivity 
to redshift distortion effects. We use perturbation theory to evaluate the magnitude of the effect on the 
volume-average skewness, for various cosmological models. One important finding of our analysis is that 
the skewness BAO signal occurs at smaller scales than that in second order statistics. For an LCDM spectrum 
with WMAP7 normalization, the BAO feature has a maximum wiggle amplitude of $~3\%$ and appears at a scale 
of $\sim82\hmpc$. We conclude that the detection of BAO wiggles in future extensive galaxy surveys via 
the skewness of the observed galaxy distribution may provide us with a useful, and potentially 
advantageous, measure of the nature of Dark Energy. 
\end{abstract}


\section{Introduction}
We live in the era of precision cosmology. The last twenty years of tremendous progress in astronomical observations 
allowed us to establish the standard cosmological model, the Lambda Cold Dark Matter model (LCDM). The cosmological 
parameters characterising the model have been determined to an accuracy of better than a percent. 

Notwithstanding the success of these measurements, we are left puzzled at finding a universe whose dynamics 
and fate are dominated by a dark energy whose identity remains a mystery. It is not even sure whether it really 
concerns an energy component to be associated to a new species in the Universe, a cosmological constant or a modification 
of gravity itself.  It is far from trivial to constrain the nature of dark energy, due to the relatively weak imprint of 
the equation of state of dark energy, in combination with sizeable observational errors \citep{frieman08}. Its principal 
influence is the way in which it affects the expansion of the Universe. In turn, this determines the growth of the 
cosmic density perturbations. 

Since the discovery of the accelerated expansion of the universe on the basis of the luminosity distance 
of supernovae \citep{riess98,perlmutter99}, a range of alternative probes have provided additional constraints 
on the nature of dark energy. The temperature perturbations in the cosmic microwave background, 
weak gravitational lensing by the large scale matter distribution and the structure formation 
growth rate are well-known examples of dark energy probes. 

The particular probe that we wish to address in this study are baryonic acoustic oscillations (BAOs), the residual 
leftover in the baryonic and dark matter mass distribution of the primordial sound waves in the photon-baryon plasma 
in the pre-recombination universe. Following the decoupling between matter and radiation, the primordial sound waves 
devolve into slight and subtle wiggles in the distribution of galaxies, the "baryonic wiggles"~\citep{wiggles1,wiggles2}. 
These have a characteristic size in the order of the sound horizon at decoupling or, more accurately, the baryon drag 
epoch at which baryons are released from the Compton drag of photons. The value of the latter is around $\sim 150 \hmpc$ 
for the standard LCDM cosmology. 

The BAO wiggles were first detected in the power spectrum of galaxies in the 2dFGRS~\citep{BAOobs1} and in the 
in the two-point correlation function of the SDSS galaxy redshift surveys~\citep{BAOdetect}. In the meantime, 
this has been followed up by a flurry of studies \citep[e.g.][]{BAOobs2,BAOobs3,BAOobs4,BAOobs5}, and have led to the 
initiation of a number of large observational program. Notable examples of surveys that (partially) probe dark energy by 
means of Baryonic Acoustic Oscillations are WiggleZ \citep{drinkwater10}, BOSS \citep{ross10} and the Dark Energy Survey, 
along with the ambitious ESA Euclid Mission project \citep{laureijs11}. 

The basic idea behind the use of BAOs as probe of dark energy is to use the sound horizon at decoupling as a standard 
ruler. Its physical size is known precisely from first principle, and its angular size was measured accurately 
in the angular spectrum of the microwave background temperature anisotropies. By assessing the size of the BAO 
characteristic scale over a range of (lower) redshifts, we may directly measure the angular diameter distance 
as a function of z, and hence constrain the nature of dark energy. The BAO scale can be inferred from the 
imprint of the sound horizon in the spatial distribution of galaxies, and can be inferred from the 
correlation function, power spectrum or other measures of spatial clustering~\citep{BAO1,BAO2,BAO3,BAO4,BAO5,BAO6}.
It was also suggested that the BAO signature could be measured with a use of higher order statistics. However this
mainly concentrated on bispectrum or three-point function alone \citep{bispec_bao1,bispec_bao2}.
Albeit, while the theoretical underpinning of BAOs is quite straightforward, the practical limitations and complications 
are considerable. A range of systematic effects, ranging from non-linearities to redshift space distortions and 
unknown galaxy biasing effects, severely complicates the analysis. 

To alleviate some of the practical complications, in this paper we advocate the idea to look for the BAO wiggles in the reduced 
third moment of the density field, the skewness $S_3$. According to perturbation theory, the $S_3$ is sensitive to second-order 
logarithmic derivative of the matter power spectrum P(k). The direct implication of this is the imprint of the BAO wiggles in 
the skewness. This forms a potentially powerful means of probing the baryonic acoustic oscillations. Unlike the more conventional 
second order probes, i.e. two-point correlation function and power spectrum, the skewness is insensitive to the bias of galaxies 
in case this involves simple linear or local bias. As important is the fact that skewness has been found to be rather marginally affected, 
in comparison to two-point correlation function or power spectrum, by redshift space distortions~\citep{s3_redshift1,s3_redshift2,s3_redshift3}.

These observations suggest that the skewness may offer a cleaner probe of the baryonic acoustic oscillations in the galaxy 
distribution, less ridden by the systematic uncertainties that still beset the present BAO experiments. In this study, we 
present the basic results for measuring the signature and scale of baryonic acoustic oscillations. We assess the BAO 
signature in the skewness of the cosmic matter density field, by invoking the theoretical power spectrum for a universe 
filled with a given baryonic fraction of matter. In \S~\ref{sec:ptbao} we present the perturbation theoretical basis 
of the use of skewness as BAO probe, including a discussion of the influence of non-linearities on the skewness 
estimates. This is followed by a presentation of the results in \S~\ref{sec:results} towards the feasibility of relating the 
skewness to the scale and amplitude of the BAO. Prospects and complications are discussed in \S~\ref{sec:discuss}. 

\section{BAO in weakly non-linear perturbation theory}
\label{sec:ptbao}
In this section we first outline the basic perturbation theory concepts underlying our BAO 
analysis, followed by a presentation of the formalism to incorporate the baryonic acoustic 
signatures in the power spectrum to evaluate the skewness $S_3$ of the field.

\subsection{Skewness and Perturbation Theory}
\label{sec:ptgen}
In our discussion on the skewness $S_3$ of the density field, we focus on the volume average second and 
third order moments. Invoking the ergodic theorem \footnote{The ergodic theorem states that the ensemble 
average of statistical quantities over a large number of different realizations of a given probability 
distribution is equal to the average of the same quantity over a sufficiently large volume, representing 
a statistically fair sample of the statistical process.} or fair-sample hypothesis, we can express the 
volume-averaged $J$-point correlation function as:
\beq
\label{volume-aver-nfunction}
\bar{\xi_J}=V_W^{-J}\int_S d\mathbf{x_1}...d\mathbf{x_J}W(\mathbf{x_1})...W(\mathbf{x_J})\xi_J(\mathbf{x_1},...,\mathbf{x_J}),
\eeq
where $\mathbf{x_i}$ is the comoving separation vector, $W(\mathbf{x})$ is a window function with volume 
\beq
V_W\,=\,\int_S d\mathbf{x}\,W(\mathbf{x})\,,
\eeq
and the integral covers the entire volume $S$. Because of the fair-sample hypothesis, $\bar{\xi_J}$ does not depend 
on the location $\mathbf{x}$ and is a function of the window volume $V_W$ only \cite{1980Peebles}\,. 

The skewness $S_3$ is the normalized ratio of third order to second order moment, defined as 
\beq
\label{eqn:s3}
S_3(R) = {\bar{\xi_3}(R)\over\bar{\xi_2}^2(R)} = {\bar{\xi_3} \over \sigma^{4}}
\eeq

The skewness of the cosmic density field has been assessed within the context of a perturbation theory analysis 
\citep{1980Peebles,Juszkiewicz1993,skew_cur_pt}. In the linear and quasi-linear regime,  \cite{Juszkiewicz1993} and 
\cite{skew_cur_pt} found that the skewness $S_3$ of the field, smoothed by a spherical top-hat window filter of 
scale $R$, is given by 
\beq
\label{eqn:s3_pt}
S_3(R) = {34\over 7} + \gamma_1(R). 
\eeq
The factor $34/7$ would have been obtained if smoothing were no taken into account, and was 
inferred by Peebles for an $\Omega=1$ cosmology. The smoothing introduces a dependence 
on scale, through the logarithmic derivative $\gamma_1(R)$ of the variance, 
\beq
\label{eqn:gamma1}
\gamma_1(R) = {d\log\sigma^2(R)\over d\log R}\;,
\eeq
where $\sigma^2(R)$ is the variance within the spherical top-hat window $W_{TH}(x)$,
\beq
\label{eqn:stdev}
\sigma^2(R) = {1\over 2\pi^2}\int_0^\infty dk\,\,\, k^2 P(k)\,W_{TH}^2(kR)\,.
\eeq
In terms of the effective slope $n_{eff}$ of the power spectrum at scale $R$, $\gamma_1(R)$ may be 
written as 
\beq
\label{eqn:gamma1n}
\gamma_1(R) = -(n_{eff}+3) \;.
\eeq
Several authors confirmed that equation~(\ref{eqn:s3_pt}) can be used as a fair estimator of 
the skewness of the top-hat filtered density field on linear and weakly non-linear scales \citep[e.g.][]{Juszkiewicz1993,skew_cur_pt,ber94,cic_ana,npoint_omega_cdm}.

\subsection{Skewness and Power Spectrum}
\label{sec:ptgen}
Perturbation theory may be used towards calculating the expected shape and amplitude of the BAO 
signal for a given cosmological model with power spectrum $P(k)$. On the basis of a few straightforward 
mathematical operations, one may infer that the $\gamma_1$ factor eqn.~(\ref{eqn:gamma1}) can be expressed in 
terms of an integral over the power spectrum $P(k)$ of the density field (we present the derivation in the appendix \ref{sec:app}):  
\beq
\label{eqn:gamma1_pk}
\gamma_1(R)=2-{2\int_0^\infty\textrm{dk}P(k)j_1(kR)kR\left[j_0(kR)-2j_2(kR)\right]\over 3\int_0^\infty\textrm{dk}P(k)j_1^2(kR)}\,,
\eeq
where $j_0$, $j_1$ and the $j_2$ are the spherical Bessel functions of the $0^{th}$, $1^{st}$ and $2^{nd}$ order. 

\bigskip
To obtain insight into the behaviour of $\gamma_1$ as a function of cosmology, we use the expressions by 
\cite{EH1,EH2} to model a linear power spectrum that incorporates BAO wiggles, 
\beq
\label{eqn:pk}
P(k) = A\cdot k^{n_s} \cdot T(k)^2\,,
\eeq
where $n_s$ is the primordial power spectrum slope, usually close to $1.0$. The transfer function $T(k)$ 
encapsulates the details of the cosmological model, including the baryonic wiggles, 
\beq
\label{eqn:tf}
T(k) = f_bT^{b}(k)+(1-f_b)T^{cdm}(k)\,,
\eeq
where $f_b\equiv{\Omega_b/\Omega_0}$ is the fraction that baryons contribute to the full matter density. The 
effects of the baryons are multifold. In addition to the relatively small acoustic oscillations, it also 
includes major effects like Silk damping. To investigate the relative importance of the baryonic oscillations, 
we compare the value of $S_3$ for power spectra based on a baryonic transfer function $T_b$ that includes 
the acoustic oscillations with those of "no-wiggles" power spectra where these signatures are averaged out. 

The amplitude $A$ of the power spectrum is normalized on the basis of the value of the (linear) $\sigma_8$ 
parameter at $z=0$, 
\beq
\label{eqn:pk_norm}
A=\sigma_8^2 \bigg/{1\over 2\pi^2}\int_0^{\infty}k^{2+n_s}T^2(k) W_{TH}(k\cdot R_{TH})\,,
\eeq
for $R_{TH}=8\hmpc$.
In our study, we have adopted the LCDM model at $z=0$ as our reference cosmology, using the 
WMAP7 best-fit values for the cosmological parameter estimates (data wmap7+bao+h0, \cite{WMAP7}):  
$\Omega_0h^2=0.134, \Omega_bh^2=0.0226, \Omega_{\lambda}=0.728, \sigma_8=0.809, n_s=0.963, h=0.704$.

\subsection{Non-linear Influences}
\label{sec:baonlin}
The linear power spectrum expression of eqn.~(\ref{eqn:pk}) will provide us with an impression of the 
influence of baryonic wiggles on the skewness $S_3$, via the integral expression of eqn.~(\ref{eqn:gamma1_pk}), 
for any given set of cosmological parameters. 

However, even on scales as large as those characteristic for BAO wiggles, $R \sim 100 \hmpc$, the 
effects of non-linear evolution of the density field become noticeable. Non-linear amplitude growth and 
mode coupling will affect the BAO signal, strongly on non-linear scales and subtly but significantly 
on quasi linear scales. A range of studies have demonstrated that non-linearities do indeed represent 
a major influence. On the basis of the analytical fitting formula by \cite{Halofit}, \cite{GuBeSm2006} showed 
that percent-level shift of the acoustic peak location in the two-point correlation function may be 
expected. Significant shifts were also found by the study of \cite{RScube06}, on the basis of perturbation theory 
and on the basis of the power spectrum of halos in simulations. Following a similar route, \cite{Angulo2008} looked 
at the impact of non-linearities on the power spectrum of galaxies in semi-analytical models. They conclude that 
percent level shifts cannot be excluded. The thorough analytical study by \cite{nonlinearBAO} has provided particularly 
strong insight into this issue, supporting the claims for a non-negligible influence. 

To account for the non-linear power spectrum evolution we have decided to use the well-known 
\verb#halofit# formula of \cite{Halofit}. We are aware of the fact that this approach may not be
the most accurate. However we can treat this procedure as a first order approximation of the non-linear effects.
The full study of the impact of the non-linearities calls for N-body experiments and will be presented
in the accompanying paper.

\subsection{BAO impact on Power Spectra}
\label{sec:baonlin}
\begin{figure}
  \includegraphics[width=70mm,angle=-90]{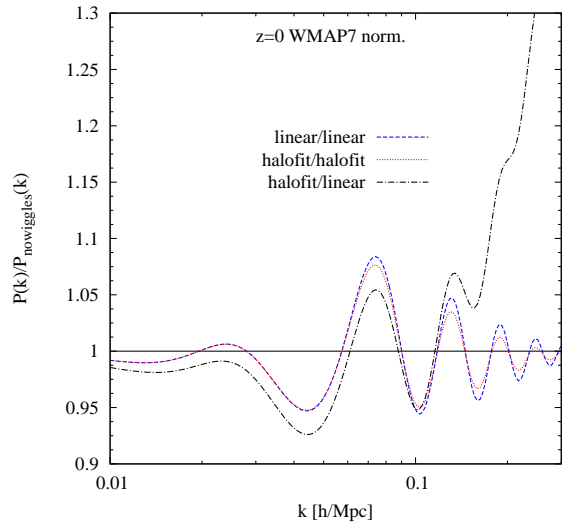}
   \caption{BAO imprint in power spectrum. To highlight the effect of the acoustic oscillations, 
we plot the ratio of the spectrum including the BAO wiggles to that of the corresponding power spectrum 
without BAO wiggles. Dashed line: ratio of linear power spectra with and without BAO. Dotted line: ratio 
of non-linear halofit corrected power spectra with and without BAO. Dashed-dotted line: ratio 
of non-linear halofit corrected power spectrum to linear ``no-wiggle" spectrum. The latter 
highlight the effects of non-linearities.}
 \label{fig1}
\end{figure}

Figure~\ref{fig1} compares the power spectra - for both linear as well as non-linear situations - that incorporate 
the BAO wiggles with those of "no-wiggle" ones where these are smoothed out. To compare these we plot the ratio of 
the spectrum with BAO wiggles to the "no-wiggle" one, so that the unity baseline represents the corresponding 
"no-wiggle" spectrum. 

We compare three pairs of spectra. The first pair compares the linear BAO wiggle spectrum to 
the corresponding linear "no-wiggle" one. For both we use the outcome from the Eisenstein \& Hu expression 
eqn.~(\ref{eqn:tf}). The second pair consists of the non-linear spectrum with BAO wiggles and the corresponding 
non-linear one with wiggles smoothed out. Both spectra are obtained from the corresponding linear spectra, 
subsequently transformed via the halofit formula. The third pair compares the full non-linear, halofit corrected, 
spectrum with baryonic wiggles to that of the no-wiggles linear spectrum. 

The first pair nicely illustrates the imprint of the baryonic acoustic oscillations. The impact of non-linearities 
is most starkly exemplified by the third pair, revealing a dramatic rise of the spectral amplitude at high frequencies. 
However, the effect of acoustic wiggles is not dramatically changed when comparing the corresponding non-linear 
spectrum to the non-linearly evolved "no-wiggle" one. Its principal effects is merely a lowering of amplitude, 
of not more than a few percent and mainly visible at higher order harmonics.

\subsection{Skewness \& Non-linearities}
\label{sec:baonlin}
On the basis of the power spectra in figure~\ref{fig1}, we may expect that non-linearities will slightly affect the 
estimates of the skewness $S_3$. The lowering of the amplitude of the BAO wiggles in the non-linear regime should 
lead to a weakening of the BAO signal in $S_3$. 

\begin{figure}
  \includegraphics[width=70mm,angle=-90]{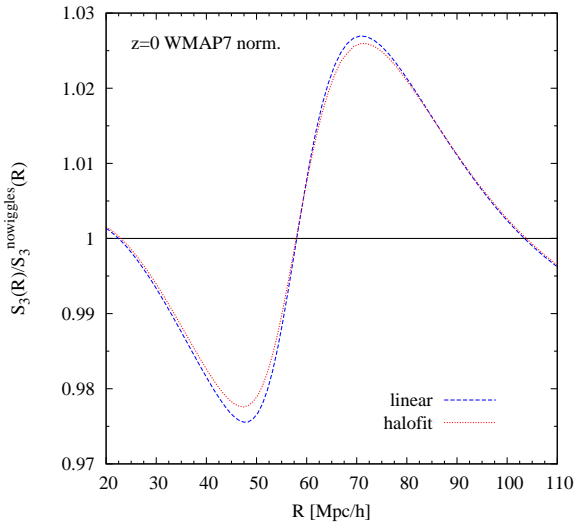}
   \caption{Reduced skewness and non-linearities. The lines depict the ratios of skewness computed using spectra including 
BAO wiggles to that computed with 'no-wiggles' spectra. Dashed line: skewness following from linear power spectra. 
Dotted line: skewness following from non-linear halofit corrected power spectra.}
 \label{fig2}
 \end{figure}

Figure~\ref{fig2} plots the skewness, computed from equations~(\ref{eqn:s3_pt}) and (\ref{eqn:gamma1_pk}), for 
linear and non-linear power spectrum. To enhance our appreciation of the effect that the BAOs have on the result, 
we plot the ratio of the skewness obtained for a power spectrum with the acoustic wiggles to that of the corresponding 
"no-wiggle" spectrum". The dashed line marks the ratio of $S_3(R)/S_3^{nowiggles}(R)$ for the linear power spectra, the dotted line 
for the non-linear halofit-corrected spectra. 

The first observation is that we indeed find a clear imprint of the baryonic oscillations in the skewness $S_3$, 
in both the linear and the non-linear situation at the 2 to 3 percent level. The amplitude of the BAO 
signal in $S_3$ is slightly smaller for the non-linear situation than the linear one. Of real 
importance is the observation that the the non-linear evolution of the power spectrum does not lead to 
a shift in the scale dependence of the skewness signal. The scale at which we find the peak and the 
minimum in the BAO impact on $S_3$ remains the same. Also, there is no change in the scale at which 
the BAO effect changes from negative to positive. As we will discuss later, the latter will be of 
crucial significance. 

\section{Results}
\label{sec:results}
 \begin{figure}
  \includegraphics[width=70mm,angle=-90]{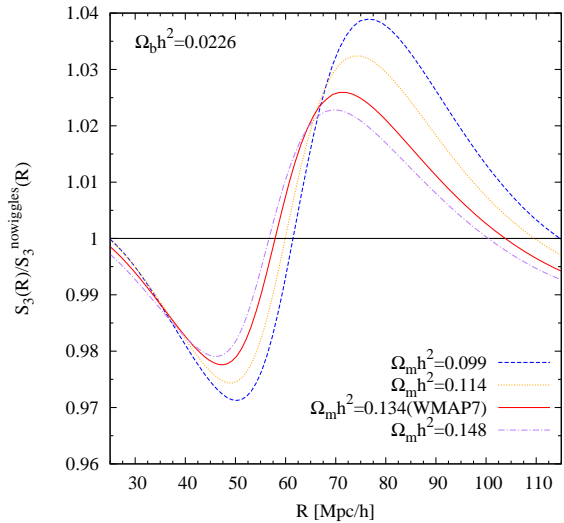}
   \caption{The acoustic oscillations imprint in the skewness $S_3$. In all cases we show the skewness computed
   using non-linear P(k) with wiggles divided by the skewness computed for 'no-wiggles' power spectrum. The lines
   depict models with different values of $\Omega_mh^2$: 0.099 - the dashed line, 0.114 - the dotted line, 
0.134 (WMAP7) - the solid line and 0.148  marked by dashed-dotted line. For all models the baryon density has
the same value $\Omega_bh^2=0.0226$. }
 \label{fig3}
 \end{figure}
Having established the basic ingredients of our study, we next turn to the principal results of this 
study. 

\subsection{Baryonic Fraction}
\label{sec:baobar}
Figure~\ref{fig3} plots the ratio of the skewness for a spectrum with and without wiggles, 
$S_3(r)/S_3^{nowiggles}(r)$, for a range of LCDM cosmologies in which baryons constitute a 
different fraction of the non-relativistic matter content of the Universe. The baryon density is the 
same for all models, $\Omega_bh^2=0.0226$, a value which is tightly constrained by the CMB temperature 
anisotropy power-spectrum (e.g. \cite{WMAP7}). The total matter density varies from 
$\Omega_mh^2=(0.099, 0.114, 0.134, 0.148)$.

In all cosmologies, the BAO skewness signal consists of a characteristic pattern. Starting 
from small scales, we observe a decrease of the skewness as the scale increases, reaching a minimum 
at scales of roughly $45-50\hmpc$. Subsequently, it climbs towards values larger than unity, which it 
reaches at a scale of $\sim 55-60\hmpc$, after which it increases towards a peak value of 
$2-4\%$ at a scale of around $75-80\hmpc$. Towards larger scales it gradually falls off towards 
unity. 

We immediately observe that the BAO skewness signal increases strongly as $\Omega_m$ is lower and 
baryons constitute a more prominent fraction of matter content of the Universe. While the amplitude 
of the BAO skewness signal is responding systematically to the cosmic baryon fraction, it does not 
seem to exceed the $\sim 3-4\%$ level. In this sense, the BAO imprint is smaller that that seen in 
the corresponding power spectrum of density perturbations, where the wiggles may amount to 
a $\sim 10\%$ effect. 

\begin{figure}
 \includegraphics[width=70mm,angle=-90]{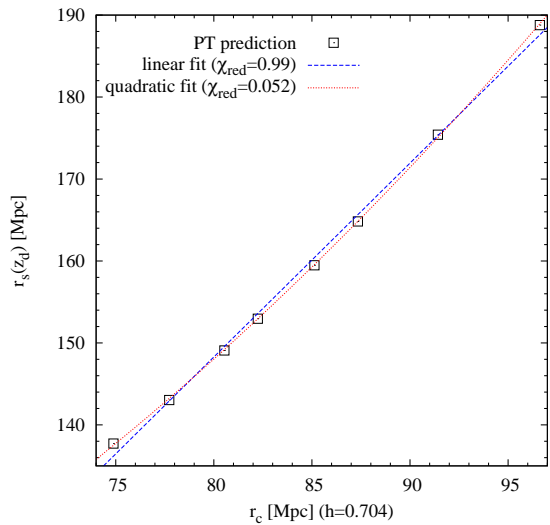}
   \caption{The dependence of the sound horizon scale ($r_s$) on the crossing scale ($r_c$). The boxes mark the values predicted
by PT skewness values (see Fig.~\ref{fig3}). Two lines show our fit to this scaling: the linear fit (the dashed line)
and quadratic fit (the dotted line). The values of $r_c$ where rescaled to Mpc using value of $h=0.704$.}
\label{fig4}
\end{figure}
\begin{table}
\caption{The sound horizon scale $r_s$ at the baryon drag epoch $z_d$ and the corresponding crossing scale ion $S_3$ signal $r_c$.
All models are computed with fixed $\Omega_bh^2=0.0226$. Values of $r_c$ are obtained with use of $h=0.704$.}
\label{tab:scales}
\begin{center}
\begin{tabular}{cccc}
\hline
$\Omega_mh^2$  & $r_s(z_d)$[Mpc] & $r_c$[Mpc] & $z_d$\\
\hline
0.099 & 164.82 & 87.24 & 1016.04\\
0.114 & 159.48 & 85.13 & 1018.14\\
0.134 & 152.95 & 82.24 & 1020.48\\
0.148 & 149.1 & 80.54 & 1021.67\\
\hline
\end{tabular}
\end{center}
\end{table}

\subsection{Skewness Scale \& Sound Horizon}
\label{sec:soundhor}
We also make the interesting observation that the main wiggle in the skewness appears at a 
considerably smaller scale than that of the BAO peak in the two-point correlation function $\xi_2(r)$ 
\citep[e.g.][]{nonlinearBAO}. This goes along with the finding that the BAO imprint on the 
skewness is visible over a considerable larger range of scales than that in the two-point 
correlation function. While for the latter, the BAO signal is noticeable on scales from roughly 
$80\hmpc$ to $130\hmpc$, the skewness is affected over a wider range of $20\hmpc$ to $100\hmpc$. 

While the amplitude change of the skewness is closely related to the effective baryon fraction $f_b$ of a cosmological model, 
the scales associated with the wiggle should be correlated with the scale of the acoustic horizon. The scale of the acoustic horizon 
$r_s$ is a fundamental aspect of the physics of baryonic acoustic oscillations. It is the physical scale of the largest 
acoustic oscillations at the epoch of recombination, and as such is the scale imprinted in the BAOs visible in the 
distribution of galaxies and baryons in the Universe. Determining the apparent BAO scale over a wide range of redshifts, 
and relating it to the sound horizon, is the goal of a large number of current and future galaxy redshift surveys.  
\citep{BAOforecast,Nbody1,BAO5,Nbody3}. 

To enable the exploitation of the information content of the skewness on the acoustic oscillations, it is therefore of 
fundamental importance to establish the connection of the sound horizon scale with the characteristic scales we find 
in the BAO feature in $S_3$. To this end, we compare the sound horizon scale $r_s$ for a range of cosmologies to 
the crossing scale $r_c$ at which the effect of BAO wiggles on the skewness reverses from negative to positive 
(see fig.~\ref{fig2} and \ref{fig3}). This scale is clearly a characteristic for the BAO imprint on skewness, 
and may be formally defined as the scale $r_c$ at which 
\beq
\Delta_{S_3}(r_c)\equiv {\displaystyle S_3(r_c) \over \displaystyle S_3^{nowiggles}(r_c)} = 1\,.
\eeq
In other words, it is the scale at which $\Delta_{S_3}(r)-1$ changes its sign from negative to positive.
We have also checked that the $r_c$ scale is corresponding with the location of the inflection point of the $S_3$ curve
near that scale.

We have measured the characteristic crossing scale for all models, and related them to the corresponding 
sound horizon scales and redshift of the baryon drag epoch - $z_d$ \citep{EH2}. We list the measured 
values in table~\ref{tab:scales}. To obtain a good understanding of the systematic relation between the 
two quantities, we have also evaluated their values for a few cosmological models that are currently 
disfavoured by observations, involving very low and very high $\Omega_m$ values. In figure~\ref{fig4} we 
have plotted the values of the sound horizon scale $r_s(z_d)$ versus the measured characteristic skewness 
crossing scale $r_c$. The figure shows that the relation is close to linear. 

In order to assess whether a higher order function would fit the relation between $r_c$ and $r_s$ better, we evaluated a 
first order and a second order fit, 

\begin{eqnarray}
\label{eqn:fit_lin}
r_s &=& a_l\cdot r_c+b_l\,\qquad\qquad\qquad\qquad \textrm(linear)\,,\\
\label{eqn:fit_quad}
r_s &=& a_q\cdot r_c^2+b_q\cdot r_c + c_q\,\qquad\qquad \textrm(quadratic)\,.
\end{eqnarray}

\medskip
\noindent For the relation shown in figure~\ref{fig4}, we have found $a_l=2.365$ and $b_l=-40.92$ for the linear fit, with 
$\chi^2_{red}=0.99$. The quadratic fit appears to be considerably better, and has a $\chi^2_{red}=0.052$ at 
parameters $a_q=0.0187$, $b_q=-0.934$ and $c_q=95.538$. 

The important conclusion from this result is that it is indeed possible to determine the sound horizon scale 
at the baryon drag epoch, $r_s(z_d)$, if we can succeed in successfully measuring the characteristic 
skewness crossing scale $r_c$.

\subsection{Skewness BAO signature}
\label{sec:skewbao}

\begin{figure}
 \includegraphics[width=60mm,angle=-90]{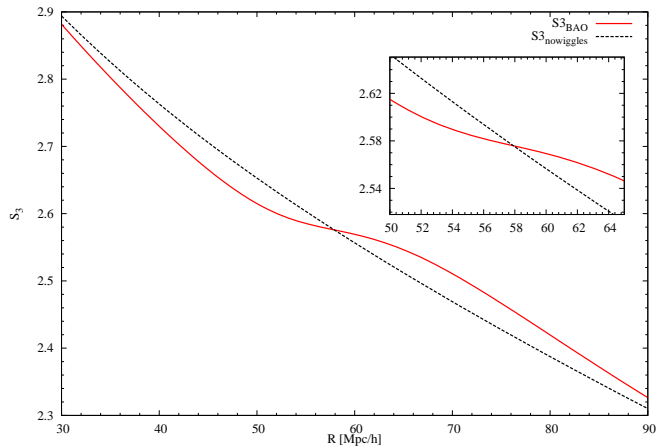}
  \caption{Skewness BAO feature. The red solid line represents the volume-averaged skewness as 
function of top-hat scale. It shows the characteristic skewness BAO feature, a mild shoulder around 
the characteristic crossing scale $r_c$. For comparison, the dotted line represents the skewness 
for a "no-wiggle" power spectrum. The insert zooms in on the shoulder at the crossing scale 
$r_c$.}
\label{fig5}
\end{figure}

As a final aspect of our analysis, in figure~\ref{fig5} we plot the behaviour of the reduced skewness 
as a function of (top-hat) scale $R$, at scales where the BAO feature is most prominent. The red solid line 
represents the skewness, as predicted by perturbation theory, for a power spectrum with BAO wiggles. 
For comparison we also plot the skewness resulting from the power spectrum with a 'no-wiggle' 
transfer function. 

For the spectrum with wiggles, we find a characteristic shoulder at around the crossing scale. 
The conclusion from this comparison is that it should indeed be possible, at the $1-2\%$ level, 
to find the imprint of baryonic acoustic oscillations. In an idealised survey, without 
systematic errors and bias, it should therefore be possible to detect the BAO feature. 

\section{Summary and Discussion}
\label{sec:discuss}
In this report we have studied the imprint of baryonic acoustic oscillations (BAOs) 
on the skewness of the density field, i.e. its reduced third moment. The skewness 
of the density field has been found earlier to be a robust test of the gravitational 
instability mechanism on cosmological scale. Amongst others, it may function as a probe 
of the nature of gravity \citep[e.g.][]{skew_gravity}, and of the initial conditions 
itself \citep{skew_nongaussian,moments_nongaussian}. This has prodded us to investigate 
whether the skewness could also help towards inferring cosmological relevant information 
via BAOs. 

To investigate the question of the BAO sensitivity of skewness, we have resorted to 
perturbation theory. From this, we obtain an integral expression for the skewness, 
dependent on the power spectrum. To model the baryonic acoustic oscillations, we have 
used the linear power spectra expressions of \cite{EH1,EH2}. To take account of the 
non-linearities in the evolving matter density field, we use the approximate analytical 
expressions of \cite{Halofit}. 

We find that the BAO skewness signal is characteristic and may not only be detectable 
in the observational reality, but may even offer a potentially powerful alternative 
to existing means of probing BAOs. The BAO skewness signal appears to have an 
amplitude in the order of $3-4\%$. Interestingly, it stretches out over a substantial 
range of scales, from $\sim 20-100\hmpc$ and occurs at much smaller scales than the 
BAO imprint on power spectrum and correlation function. Perhaps most importantly, 
we have established a strict, near-linear relation between the sound horizon scale at 
the baryon drag epoch and the scale at which the BAO skewness signal crosses from 
negative to positive. 

Even though the amplitude of the BAO skewness signal is somewhat lower than 
that of the BAO wiggles in the power spectrum and correlation function, it is far less 
sensitive to several systematic effects that still represent a major challenge for 
inferring cosmological parameters on the basis of BAO measurements in observational surveys. 
One major advantage of the BAO skewness signal over that in second order measures 
like the two-point correlation function and power spectrum, is that it is less 
sensitive to the bias of the galaxy population with respect to the mass distribution. 
If the galaxy density is a local function of the mass density, the relation between 
the skewness and the variance of the density field is preserved \citep{s3_bias1,s3_bias2}. 
In other words, the shape of the $S_3$ reduced skewness as a function of scale is preserved
when one concerns linear and local biasing of the density field. 
A second, and perhaps even more 
prominent advantage, is that the skewness - in the weakly non-linear regime - is relatively 
insensitive to redshift space distortions~\citep{s3_redshift1,s3_redshift2,s3_redshift3}. 

While this short publication is meant to establish the feasibility of using skewness 
to probe BAOs, there are numerous issues and details that will be discussed and evaluated 
in an upcoming study. One major issue is the influence of non-linearities in the 
density field. Subtle non-linear effects may go beyond the scope of what the non-linear 
halofit power spectra may model. Also, the complexities of galaxy bias may only be 
adequately modelled by means of large N-body simulations and realistic galaxy formation 
models. On the other hand, the current observational estimates of $S_3$ from the SDSS \citep{skew_obs_sdss}
and 2dFGRS \citep{skew_obs_2df1,skew_obs_2df2} are characterised by relative big errors. This suggests
that detection of the BAO feature will be difficult. In the end, proper modelling of hydrodynamical and radiative processes will be 
essential for a truly full treatment of non-linear and biasing effects \citep{skewness_barions}. 

\section*{Acknowledgements}
Soon after we started work on this project Roman Juszkiewicz had undergone a dramatic deterioration 
of health, as a result of which he passed away on 28th January 2012.  We leave this paper as a 
final tribute to a leading scientist and teacher who was a great friend to his entire community.

The authors would like to thank Carlton Baugh, Elise Jennings, Bernard Jones and Changbom Park for 
useful discussions. WAH acknowledge  the support of this research received from Polish National
Science Center in grant no. DEC-2011/01/D/ST9/01960 and from the Institute of Astronomy of the University of Zielona G\'ora.

\bibliographystyle{mn2e_fixed}
\bibliography{bao_skewness}

\bsp

\appendix
\section{Derivation of the analytical form for the $\gamma_1(R)$}
\label{sec:app}
To derive expression~\ref{eqn:gamma1_pk} for the $\gamma_1$ parameter, we proceed as follows. The 
$\gamma_1$ parameter is defined as 
\beq
\gamma_1\equiv -{{\difd \log\sigma^2(R)}\over{\difd\log R}}\,,
\label{eqn:gam_def}
\eeq
with $\sigma^2(R)\equiv\langle\delta^2(R)\rangle$ the second moment of the density contrast 
smoothed on (comoving) scale $R$. For a tophat filter, the latter is defined as
\beq
\label{eqn:de_smooth}
\delta_R(\vec{x})=\int\delta(\vec{x'})W_R(|\vec{x}-\vec{x'}|)\difd^3x'\,, 
\eeq
where the expression for the tophat filter $W_R(x)$ is 
\beq
\label{eqn:tophat_def}
W_R(x)\equiv\left\{\begin{array}{ll}
                    {3/ 4\pi R^3}\,,\quad x\leq R\,,\\
                     0,\quad\quad\quad\quad x> R\,.
                   \end{array}\right.
\eeq
Using the integral expression for the density contrast, we find the following integral expression 
for the second moment of the density field (\cite{1980Peebles}):
\beqa
\langle\delta_R^2(\vec{x})\rangle = \nonumber\\
\iint\difd^3x'\difd^3x''\langle\delta(\vec{x'})\delta(\vec{x''})\rangle W_R(|\vec{x}-\vec{x'}|)W_R(|\vec{x}-\vec{x''}|)\nonumber\\
=\iint\difd^3x'\difd^3x''\xi(\vec{x'}-\vec{x''})W_R(\vec{x'})\cdot W_R(\vec{x''})\,.\label{eqn:de2}
\eeqa

\medskip
\noindent It is convenient to evaluate this integral expression in Fourier space. To this end, we use the following 
Fourier conventions, 
\beqa
\label{eqn:de-x}
\delta(\vec{x})&=&(2\pi)^{-3/2}\int\delta_k e^{i\vec{k}\cdot\vec{x}}\difd^3k\,,\nonumber\\
\delta_k&=&(2\pi)^{-3/2}\int\delta(\vec{x})e^{-i\vec{k}\cdot{x}}\difd^3x\,.
\eeqa
Recasting the double integral in equation~\ref{eqn:de2} to Fourier space yields
\beq
\label{eqn:sigma2R}
\sigma^2(R)=(2\pi)^{-3}\iint\difd^3k\difd^3k'\langle\delta_k\delta_{k'}\rangle e^{i\vec{k}\cdot\vec{x}+i\vec{k'}\cdot\vec{x}}W(kR)W(k'R)\,,
\eeq
leading to the following expression in terms of the power spectrum $P(k)$,
\beqa
\label{eqn:sigma2R-2}
\sigma^2(R)&=&(2\pi)^{-3}\int\difd^3kP(k)W^2(kR) \nonumber\\
&=& {1\over2\pi^2}\int_0^{\infty}\difd k k^2P(k)W^2(kR)\,.
\eeqa
where the power spectrum is defined as the Fourier transform of the variance, 
\beq
\label{eqn:sig-pk}
\langle\delta_k\delta_{k'}\rangle = P(k)\delta_D(\vec{k}+\vec{k'})\,,
\eeq
in which $\delta_D$ is the Dirac delta function. The Fourier expression for the 
tophat filter, $W(kR)$, is given by 
\beq
\label{eqn:5}
W(kR) ={3\over(kR)^3}(\sin kR - kR\cos kR)={3\over x}j_1(x)\,,
\eeq
with $j_1(x)$ the first order spherical Bessel function. 

\bigskip
\noindent Differentiation of $\sigma^2(R)$ by $R$ yields 
\beq
{\difd\sigma^2_R\over\difd R}={\difd\sigma^2_R\over\difd R}={1\over\pi^2}\int_0^\infty\difd kP(k)k^3W(kR){\displaystyle\difd W(kR)\over
\displaystyle\difd(kR)}\,,
\eeq
where we have used the fact that
\beq
{\difd W^2(kR)\over\difd R}= 2kW(kR){\difd W(kR)\over\difd(kR)}\,.
\eeq
From this, we may immediately find the expression for $\gamma_1$, 
\beqa
\gamma_1&=& -{{\difd \log\sigma^2(R)}\over{\difd\log R}}\nonumber\\
&=&-{2\int_0^\infty\difd kP(k)k^2(kR)W(kR){\displaystyle\difd W(kR)\over \displaystyle\difd(kR)}\over\int_0^\infty\difd kP(k)k^2W^2(kR)}\,.
\label{eqn:gamma1w}
\eeqa

\bigskip
\noindent Recalling the fact that the spherical tophat filter is directly related to the first order Bessel function, $W(x)={3\over x}j_1(x)$, 
and using the derivation relations for Bessel functions, we find for the derivative of the tophat function, 
\beq
\label{eqn:der-W}
{\difd W\over\difd x}= - {3\over x^2}j_1(x)-{3\over x}j_1'(x)=-{3\over x^2}j_1(x)+{1\over x}\left[j_0(x)-2j_2(x))\right]\,.
\eeq
and inserting this relation into equation~\ref{eqn:gamma1w}, after some manipulation immediately leads to the following 
expression, 
\beq
\label{eqn:13}
\gamma_1(R)=2-{2\int_0^\infty\difd kP(k)j_1(kR)kR\left[j_0(kR)-2j_2(kR)\right]\over 3\int_0^\infty\difd kP(k)j_1^2(kR)}\,.
\eeq
This relation for $\gamma_1$ is the central expression of this paper.

\label{lastpage}

\end{document}